\documentstyle[twocolumn,aps,psfig]{revtex}
\begin{document}

\twocolumn[\hsize\textwidth\columnwidth\hsize\csname @twocolumnfalse\endcsname

\title{Phase-field approach for faceted solidification}

\author{Jean-Marc Debierre$^1$, Alain Karma$^2$, Franck Celestini$^3$
and Rahma Gu\'erin$^1$}

\address {$^1$Laboratoire Mat\'eriaux et Micro\'electronique
de Provence (UMR 6137),
Universit\'e d'Aix-marseille III,
Facult\'e des Sciences et Techniques de Saint-J\'er\^ome,
Case 151,
13397 Marseille Cedex 20, FRANCE}

\address{$^2$Department of Physics and Center for Interdisciplinary 
Research on Complex systems,
Northeastern University, Boston, MA 02115, U.S.A.}

\address {$^3$Laboratoire de Physique de la Mati\`ere condens\'ee
(UMR 6622),
Universit\'e de Nice-Sophia-Antipolis,
Parc Valrose,
06108 Nice Cedex 2, FRANCE}

\maketitle

\vskip1pc

\begin{abstract}
{
We extend the phase-field approach to model the solidification of faceted
materials. Our approach
consists of using an approximate $\gamma$-plot with rounded
cusps that can approach arbitrarily closely the true $\gamma$-plot
with sharp cusps that correspond to faceted orientations.
The phase-field equations are solved in the thin-interface limit
with local equilibrium at the solid-liquid interface
[A. Karma and W.-J. Rappel, Phys. Rev. E{\bf 53}, R3017 (1996)]. The 
convergence
of our approach is first demonstrated for equilibrium shapes.
The growth of faceted needle crystals in an
undercooled melt is then studied as a function of undercooling and the
cusp amplitude $\delta$ for a $\gamma$-plot of the form
$\gamma=\gamma_0[1+\delta(|\sin\theta|+|\cos \theta|)]$. The
phase-field results are consistent with the scaling law
$\Lambda\sim V^{-1/2}$ observed experimentally, where
$\Lambda$ is the facet length and $V$ is the growth rate.
In addition, the variation of $V$ and $\Lambda$ with $\delta$ is
found to be reasonably well predicted by an approximate
sharp-interface analytical theory that includes capillary effects
and assumes circular and parabolic
forms for the front and trailing rough parts of the needle crystal, 
respectively.}
\end{abstract}


\vskip2pc]

\section{Introduction}

Over the last decade, the phase-field approach \cite{Langer,Collins&Levine}
has been developed  extensively to model the solidification of both pure
materials and alloys \cite{Reviews}. Most of the work to date
has focused on the case where the excess free-energy of the
solid-liquid interface
\begin{equation}
\gamma\equiv \gamma_0f(\theta)
\end{equation}
is a smooth function
of the angle $\theta$ between the direction
normal to the interface and some fixed crystalline axis. In particular,
the simple form
\begin{equation}
f(\theta)=1+\epsilon\cos 4\theta+\dots,\label{weak}
\end{equation}
appropriate for a weakly anisotropic material with an underlying cubic symmetry
has been widely used in studies of dendritic solidification
\cite{Wheetal93,WanSek96,KarRap,Proetal,PlaKar,Kar01}.
For a smooth $\gamma$-plot, the value of the
diffusion field (dimensionless undercooling
or supersaturation) at the interface is given by the standard
Gibbs-Thomson condition
\begin{equation}
u=-d_0\left[f+\frac{d^2f}{d\theta^2}\right]\kappa,\label{gt}
\end{equation}
where $d_0$ is a microscopic capillary length (thermal or chemical)
that is proportional to $\gamma_0$ and
$\kappa$ is the interface curvature. Dendritic growth in pure materials
\cite{KarRap,Proetal,PlaKar} and alloys \cite{Kar01} has been
modeled quantitatively with a thin-interface limit \cite{KarRap} of 
the phase-field
model that yields the boundary condition
(\ref{gt}) and overcomes the stringent computational constraint
associated with a finite interface thickness. These simulation
studies modeled the common situation of a weak anisotropy
where the interface stiffness $\gamma+d^2\gamma/d\theta^2$,
and hence the square bracket in Eq.~(\ref{gt}),
is everywhere positive. More recently, the phase-field approach has
also been successfully extended to model strongly anisotropic
equilibrium crystal shapes where the stiffness becomes
negative for some range of $\theta$ \cite{JJ}. In this case, which occurs for example
when $\epsilon>1/15$ for the form (\ref{weak}) of $f(\theta)$, the slope
of the interface has discontinuities that must be present
to eliminate thermodynamically
unstable, and hence forbidden, orientations for which the
stiffness is negative.

In this paper, we extend the phase-field approach to model a wide
class of materials that form facets for a discrete set of
orientations. The interfacial energy is generally non-analytic for orientations
close to a facet. This non-analyticity is reflected in the presence
of cusps in the $\gamma$-plot that are of the form
\begin{equation}
f(\theta)\approx 1+\delta\,|\theta-\theta_c|+\dots~~~~{\rm for}~~~~
|\theta-\theta_c|\ll 1, \label{nearcusp}
\end{equation}
where $\theta=\theta_c$ is the orientation of a given facet.
In the simplest model where vicinal surfaces are assumed to consist
of straight ledges of height $a$ separated by terraces of width
$l\approx a/|\theta-\theta_c|$, the cusp amplitude 
$\delta=\gamma_L/(a\gamma_0)$
where $\gamma_L$ is the ledge energy per unit length.

To a good first approximation, the tendency of a material to facet is
described by Jackson's well-known
$\alpha$-factor \cite{Jac58} that is the product of a
dimensionless crystallographic factor that depends on the crystal
structure and the orientation of the interface and the ratio of the latent heat
per mole to the rare gas constant.
Jackson's theory predicts that materials with
$\alpha \le 2$ (such as metals with low entropy of melting) will
have rough interfaces, while those with
$\alpha \ge 2$ (such as non-metals and compounds) will form facets, 
consistent with
experimental observations across a wide range of materials \cite{Woo}.

The growth of faceted dendrites has been studied both
experimentally \cite{Franck,Maurer} and theoretically
in the context of a sharp-interface approach \cite{BenPom88,Vincent,Mokhtar}.
Experimentally, facets are seen to appear
near the tip of the needle crystal that nevertheless retains a 
parabolic shape on
the larger scale of the whole crystal that
includes the trailing rough parts. From a theoretical standpoint,
a main difficulty to model faceted growth is that the standard
Gibbs-Thomson relation can only be applied to rough parts
of the interface.
On a facet, this relation takes the
form of an integral condition that is obtained by
making the substitution $\kappa=d \theta/ds$,
where $s$ denotes the arclength along the interface,
and integrating both sides of Eq.~(\ref{gt})
from one extremity ($s_-$) of the facet to the other ($s_+$).
This integration yields the condition
\begin{equation}
   \int_{s_-}^{s^+} u\, ds = -d_0
   \left(\left.\frac{df}{d\theta}\right|_+
-\left.\frac{df}{d\theta}\right|_- \right)= -2\,d_0\,\delta
,\label{int}
\end{equation}
where $df/d\theta|_\pm$ denotes the limiting values of
$df/d\theta$ on each side of the cusp
($\theta-\theta_c\rightarrow 0^\pm$).

For an equilibrium crystal with a constant undercooling
or supersaturation $(u<0)$ along
the interface, the above condition implies
that the length $\Lambda$ of
the facet is simply proportional to the amplitude
of the cusp
\begin{equation}
\Lambda = 2\frac{d_0}{-u}\,\delta.
\end{equation}
For growth outside of equilibrium, the determination of the
facet length and the shape of the rough parts
requires in general a self-consistent solution of the free-boundary
problem defined by the appropriate diffusion equations for $u$ in
solid and liquid, the standard Stefan condition of heat or mass
conservation on the interface, and the local equilibrium
conditions (\ref{gt}) and (\ref{int}) on rough parts and
facets, respectively. We defer a discussion      
of kinetic effects to the conclusion section of this paper.
 
An analytical solution to this problem was obtained
for steady-state growth by Adda-Bedia and Hakim \cite{Vincent}
in the small P\'eclet number limit neglecting capillary effects
on the rough parts. They concluded from this analysis that it
is not possible to require both tangential matching of the rough
and faceted parts of the interface and the
equilibrium condition ($u\approx 0$) on the front and
trailing rough parts. Furthermore, they proposed an approximate
solution to the full problem that includes capillary effects on the
rough parts in the large $\delta$ limit. In this solution,
the facets are assumed to match tangentially to a small
quasi-circular tip that is significantly more
undercooled than the rough trailing parts. A numerical solution
of the steady-state growth problem was later obtained
for arbitrary $\delta$ by Adda-Bedia and Ben Amar \cite{Mokhtar}
using a boundary integral method. This calculation gave results
that are consistent with the the approximate solution of
Adda-Bedia and Hakim for large $\delta$.

At present, a phase-field approach for faceted growth would be
highly desirable to study the stability of needle crystal solutions as
well as to explore the full dynamical range of morphological evolution
when the interface shape is non-stationary.
Treating separately the rough and faceted parts
of the interface appears to be difficult within
a phase-field formulation. To circumvent this difficulty,
we follow a procedure that is similar in spirit, but different
in details, to the one introduced
by Adda-Bedia and Ben Amar \cite{Mokhtar}
in a sharp-interface context. The basic idea
is to use a regularized form of the
$\gamma$-plot that can approximate
arbitrarily closely the $\gamma$-plot with sharp cusps.
The discontinuity of $df/d\theta$ at a cusp can be viewed
mathematically as the stiffness behaving as a sharply peaked
Dirac delta function at $\theta=\theta_c$. Therefore, different
regularizations of this discontinuity can be viewed as
different regularizations of this delta function peak
that must satisfy the cusp condition
\begin{equation}
\lim_{\theta_0\rightarrow 0} \int_{\theta_c-\theta_0}^{\theta_c+\theta_0}
d\theta \left[f+\frac{d^2f}{d\theta^2}\right]=2\delta,
\end{equation}
imposed by the form (\ref{nearcusp}) of $f$ near a cusp.
Adda-Bedia and Ben Amar
have introduced regularizations in which all derivatives of
the stiffness and hence $f$ are continuous. In a phase-field
context, we have found
more convenient to use a regularization where only
$f$ and its first derivative are continuous and
the dimensionless stiffness $f+d^2f/d\theta^2$ is a step function
regularization of the delta function,
$f+d^2f/d\theta^2\approx \delta/\theta_0$ for
$\theta_c-\theta_0\le \theta \le \theta_c+\theta_0$.
In practice, computations for small $\theta_0$
turn out to be more costly because the larger stiffness
($\approx \delta/\theta_0$) requires a smaller time step.
As we shall see, however, an accurate extrapolation to
the $\theta_0\rightarrow 0$ limit is nevertheless possible.

The phase-field model and the details of this smoothing procedure
are described in the next section. The numerical implementation of
the equations is then presented in section III. The convergence of our
numerical approach is demonstrated in section IV by a detailed comparison
of diffuse-interface and sharp-interface equilibrium crystal shapes.
Results concerning the steady-state growth
of faceted needle crystals in a pure undercooled
melt are then presented in section V.
These results are then compared in section VI to the predictions
of an approximate sharp-interface theory that assumes
simple forms for the front and trailing rough parts as an extension
of the solution proposed in Ref. \cite{Vincent}. Concluding
remarks and future prospects are then discussed in section VI.

\section{Phase-field Model}

We consider the growth of a crystal
from a pure undercooled melt with the thermal
diffusivity being the same in the solid and liquid.
We follow the phase-field approach developed
by Karma and Rappel \cite{KarRap} that has proven
successful to model efficiently non-faceted dendritic growth.
We briefly recall the basic features of the phase-field
model and of the thin-interface limit used to relate
the phase-field and sharp-interface models.

The dimensionless temperature field is defined as $u=(T-T_M)/(L/c_p)$,
where $T$ is the temperature, $T_M$ the melting temperature, $L$
the latent heat per unit volume, and $c_p$ the specific heat at constant
pressure per unit volume.
The phase field $\psi$, is taken to be
equal to $+1$ ($-1$) in the solid (liquid) phase, and
varies continuously across the diffuse interface. In two dimensions,
the time evolution of both fields is governed by
the equations
\begin{equation}
\partial_t u=D\nabla^2u+{1\over 2} \partial_t\psi, \label{PF1}
\end{equation}
and
\begin{eqnarray}
\tau(\theta)\partial_t\psi & = & \big[ \psi-\lambda u(1-\psi^2)\big] 
(1-\psi^2)  \nonumber \\
& & +\vec \nabla\cdot\big[ W(\theta)^2\vec \nabla \psi\big]
-\partial_x\big[ W(\theta) W'(\theta) \partial_y \psi \big] \nonumber \\
& & +\partial_y\big[ W(\theta) W'(\theta) \partial_x \psi \big], \label{PF2}
\end{eqnarray}
where $D$ is the thermal diffusivity, $\theta$
is the angle between the normal to the interface and the $x$ axis,
$\lambda$ is a dimensionless coupling constant
between the temperature and phase fields, and
\begin{equation}
W(\theta)\equiv W_0 f(\theta), \label{Wtheta}
\end{equation}
is the diffuse interface thickness.
In the thin-interface limit where $W_0$
is much smaller than the mesoscale of the growing crystal \cite{KarRap},
the above phase-field equations reduce to the standard sharp-interface
model of diffusion-limited growth with the velocity-dependent form
of the Gibbs-Thomson condition
\begin{equation}
u=-d_0\left[f(\theta)+\frac{d^2f(\theta)}{d\theta^2}\right]\kappa
-\beta(\theta)v_n,\label{gtvel}
\end{equation}
where
\begin{equation}
d_0={a_1 W_0\over \lambda},
\end{equation}
and
\begin{equation}
\beta(\theta)={a_1 \over \lambda}{\tau(\theta) \over W(\theta)}
\Big[ 1-a_2 \lambda{W(\theta)^2 \over D\tau(\theta)}\Big].
\end{equation}
In addition, $a_1$ and $a_2$ are constants that depend generally on
the choice of the double-well potential and other functions in
the phase-field model. For the present choices that are the same
as in Refs. \cite{KarRap}, $a_1=0.8839\dots$
and $a_2=0.6267\dots$.

To model faceted growth, we
use the simplest form of the $\gamma$-plot
\begin{equation}
f(\theta)=1+\delta\big(\vert \sin \theta \vert +
\vert \cos \theta \vert\big).
\end{equation}
This form is directly adapted from the
broken-bond model which describes satisfactorily
faceted solid-gas interfaces \cite{Porter}. The same form
was used in previous sharp-interface
calculations \cite{Vincent,Mokhtar}.

Furthermore, in order to model the limit with local
equilibrium at the interface, i.e vanishing interface
kinetics $\beta(\theta)=0$, we impose \cite{KarRap}
\begin{equation}
\tau(\theta)=\tau_0 f(\theta)^2,
\end{equation}
and
\begin{equation}
\lambda ={1 \over a_2} {D\tau_0 \over W_0^2},
\end{equation}
which makes $\beta(\theta)$ vanish for
all values of $\theta$.

\begin{figure}
\centerline{
\psfig{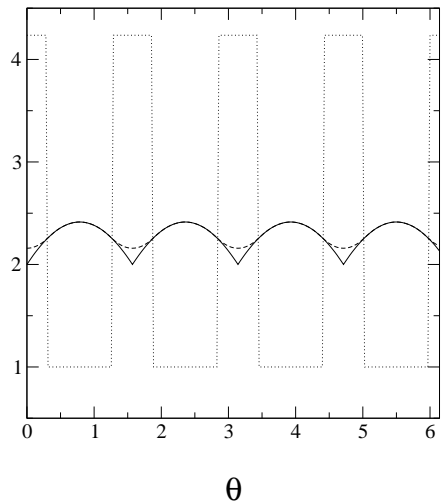}}
\smallskip
\caption{
Three functions of the interface orientation
angle $\theta$ represented for a cusp amplitude $\delta=1.0$.
Solid line: anisotropy function with sharp cusps, $f(\theta)$.
Dashed line: smoothed anisotropy function, $f_s(\theta)$.
Dotted line: dimensionless stiffness, $f_s(\theta)+f_s''(\theta)$.
A large smoothing angle, $\theta_0=\pi/10$, is
used here to make the difference between $f$ and $f_s$ visible.
}
\label{fig:ftheta}
\end{figure}

For the above choice of $\gamma$-plot,
the dimensionless stiffness defined by
\begin{equation}
S(\theta)\equiv f(\theta)+\frac{d^2f(\theta)}{d\theta^2}
\end{equation}
is constant except at the cusps,
$\theta= \theta_{c,n} = n \pi/2 $ (with $n$ an integer),
where it diverges. The discontinuities in the first derivative of
the interfacial energy need to be regularized, since $f'(\theta)$
is not defined for $\theta=\theta_{c,n}$. A simple strategy
to circumvent this problem is to smooth out
the cusps by replacing $f(\theta)$ with a smooth function
like a sine in a small range of $\theta$ values around the cusps
(Fig.~\ref{fig:ftheta}). For the sake of clarity, we restrict our discussion
to the first quadrant, $\theta \in [0,{\pi\over 2}]$, since
the problem has a natural four-fold symmetry.
The smoothed anisotropy function $f_s(\theta)$ reads then:
\begin{equation}
f_s(\theta) = \left\{
\begin{array}{ll}
1+\delta\big(\sin \theta + \cos \theta \big) &
\quad {\rm if}\quad \theta_0<\theta<{\pi\over 2}-\theta_0, \\
B-A \cos \theta  & \quad {\rm if}\quad \theta\le\theta_0,\\
B-A \sin \theta  & \quad {\rm if}\quad \theta\ge{\pi\over 
2}-\theta_0,  \label{Fs}
\end{array}
\right.
\end{equation}
and its derivative
\begin{equation}
f'_s(\theta) = \left\{
\begin{array}{ll}
\delta\big(\cos \theta - \sin \theta \big) & \quad {\rm if}\quad 
\theta_0<\theta<{\pi\over 2}-\theta_0,  \\
A \sin \theta & \quad {\rm if}\quad \theta\le\theta_0, \\
-A \cos \theta & \quad {\rm if}\quad \theta\ge{\pi\over 2}-\theta_0. 
\label{F's}
\end{array}
\right.
\end{equation}
The two constants $A$ and $B$ are obtained by expressing
the continuity of $f_s$ and $f'_s$ at $\theta=\theta_0$,
\begin{equation}
A=\delta(\cot \theta_0 -1),
\end{equation}
and
\begin{equation}
B=1+\delta / \sin \theta_0.
\end{equation}
Now $f'_s(0)=f'_s(\pi/2)=0$ and $W(\theta)$ is no more singular
in the cusp directions. Note, however, that $f''_s$ is {\em not}
continuous at $\theta=\theta_0$. Consequently, the dimensionless stiffness is
a step function of $\theta$ (see Fig.~\ref{fig:ftheta}),
\begin{equation}
S(\theta) = \left\{
\begin{array}{ll}
1 & \quad {\rm if}\quad \theta_0<\theta<{\pi\over 2}-\theta_0, \\
1+\delta / \sin \theta_0 & \quad {\rm if}\quad \theta\le\theta_0
{\hskip10pt\rm or\hskip10pt}\theta\ge{\pi\over 2}-\theta_0.
\end{array}
\right.
\end{equation}

\section{Numerical implementation}

We now briefly describe how Eqs.~(\ref{PF1}, \ref{PF2}) are 
discretized in our code.
The interface is represented by the $\psi=0$ contour, so that
$\vec \nabla \psi$ is by construction colinear to the unit vector along
the normal to the interface,
{\bf n}. If $\vert \vec \nabla \psi \vert \ne 0$, the two components
of this vector are given by
\begin{equation}
n_x=\cos \theta =- \partial_x \psi / \vert \vec \nabla \psi \vert,
\end{equation}
and
\begin{equation}
n_y=\sin \theta =- \partial_y \psi / \vert \vec \nabla \psi \vert,
\end{equation}
and $f_s$ and $f'_s$ are computed according to Eqs (\ref{Fs}, \ref{F's}).
Conversely, when $\vert \vec \nabla \psi \vert = 0$, one sets
\begin{equation}
f_s(\theta)=1,
\end{equation}
and
\begin{equation}
f'_s(\theta)=0.
\end{equation}
Numerical simulations of Eqs.~(\ref{PF1}, \ref{PF2}) are performed
by implementing a finite-difference scheme on a regular square mesh
with mesh size $h=0.4 W_0$.
The domain considered is the quadrant $0\le x\le Nh$ and $0\le y \le Nh$,
with $N$ an integer.
A standard first-order in time Euler scheme
with a time step $\Delta t$ and a second-order in space discretization
of the spatial derivatives are used.

Let $u_{i,j}$ and $\psi_{i,j}$ respectively denote the discretized 
reduced temperature
and phase field at point $(x=ih,y=jh)$.
In a first step, one computes
$\nabla^2 u$, $\nabla^2 \psi$, $\partial_x \psi$, $\partial_y \psi$,
$W(\theta)W'(\theta)$ and $W(\theta)$ for each point of the extended domain,
$i$ and $j \in [-1,N+1]$, and the results are stored in six 
intermediate arrays.
Reflecting conditions are imposed on $u$ and $\psi$ at all the domain 
boundaries
($i$ or $j=-2,-1,N+1,N+2$), and using centered differencing 
approximations, we have
\begin{equation}
\nabla^2 u_{i,j}=\big[u_{i+1,j}+u_{i-1,j}+u_{i,j+1}+u_{i,j-1}
-4u_{i,j}\big]/h^2,
\end{equation}
\begin{equation}
\nabla^2 \psi_{i,j}=\big[\psi_{i+1,j}+\psi_{i-1,j}+\psi_{i,j+1}+\psi_{i,j-1}
-4\psi_{i,j}\big]/h^2,
\end{equation}
\begin{equation}
\partial_x \psi_{i,j}=\big[\psi_{i+1,j}-\psi_{i-1,j}\big]/(2h),
\end{equation}
\begin{equation}
\partial_y \psi_{i,j}=\big[\psi_{i,j+1}-\psi_{i,j-1}\big]/(2h).
\end{equation}
The two remaining arrays, $(WW')_{i,j}$ and $W_{i,j}$, are given by 
Eq.~(\ref{Wtheta})
into which $f$ is replaced with $f_s$.

In the second step, one solves Eqs.~(\ref{PF1}) and (\ref{PF2}) on the inner
points of the domain, $i$ and $j \in [0,N]$.
To do so, the last three terms in Eq.~(\ref{PF2}) are written
in a slightly different form,
\begin{eqnarray}
&&W(\theta)^2 \nabla^2 \psi+2W(\theta)\vec\nabla W(\theta) \vec\nabla\psi\\
&&-(\partial_y\psi)\partial_x\big[W(\theta) W'(\theta)\big]\nonumber
+(\partial_x\psi)\partial_y\big[W(\theta) W'(\theta)\big],
\end{eqnarray}
so that each new term can be computed with the help
of the intermediate arrays. For instance the last term is discretized as
\begin{equation}
(\partial_x \psi_{i,j})\big[(WW')_{i,j+1}-(WW')_{i,j-1}\big]\big/ (2h).
\end{equation}

\section{Equilibrium shapes}

\subsection{Comparison between analytical and phase-field results}
We first check our code by computing the equilibrium shape
of the crystal.
In thermal equilibrium, $u=-\Delta$ everywhere, and the evolution equation
for the phase field reduces to
\begin{eqnarray}
\tau(\theta)\partial_t\psi &=&
\big[\psi+\lambda \Delta(1-\psi^2)\big](1-\psi^2) +W(\theta)^2 \nabla^2 \psi \nonumber\\
&&+2W(\theta)\vec\nabla W(\theta) \vec\nabla\psi 
-(\partial_y\psi)\partial_x\big[W(\theta) W'(\theta)\big]  \nonumber\\
&&+(\partial_x\psi)\partial_y\big[W(\theta) W'(\theta)\big]. \label{EqPF}
\end{eqnarray}
Initially, the solid is given a circular (or a square) shape and
the undercooling is set to $\Delta=\Delta_0$.
Time integration of Eq.~(\ref{EqPF}) is performed and the interface 
velocity along
the $x$ axis, $V_x$ is computed at regular time intervals.
If $V_x>0$, $\Delta$ is decreased and it is increased if $V_x<0$.
The increment $\delta\Delta$ is divided by a constant $a>1$
each time $V_x$ changes sign.
This procedure is repeated as long as $\delta\Delta$ is larger
than some prescribed value, $\delta\Delta_{min}$. This scheme
is known to converge to an equilibrium state \cite{KarRap}.

\begin{figure}
\centerline{
\psfig{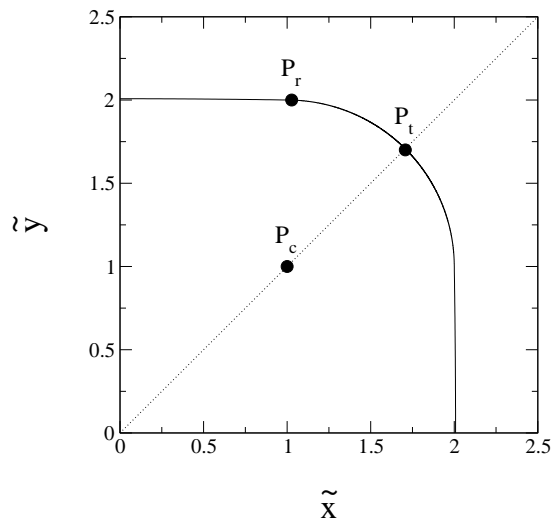}}
\smallskip
\caption{
Analytical equilibrium shape for anisotropy $\delta=1.0$
and smoothing angle $\theta_0=\pi/200$. $P_r$ is the rightmost
point of the ``facet'', $P_t$ the intersection point with
the $x=y$ line, and $P_c$ the center of the arc circle 
$\stackrel\frown {P_rP_t}$.
}
\label{fig:analeq}
\end{figure}

The analytical equilibrium shape is given by
\begin{equation}
\left\{
\begin{array}{ll}
\tilde x= x(\theta)/R_0=f_s(\theta) \cos \theta-f'_s(\theta) \sin \theta,\\
\tilde y= y(\theta)/R_0=f_s(\theta) \sin \theta+f'_s(\theta) \cos \theta,\\
\end{array}
\right.
\end{equation}
where $R_0=d_0/\Delta$ \cite{Vorhees}.
Let us consider one eighth of this interface
corresponding to $\theta \in [\pi/4,\pi/2]$
(Fig.~\ref{fig:analeq}).
In the cusp region, $\theta \in [\pi/2-\theta_0,\pi/2]$, one has
\begin{equation}
\left\{
\begin{array}{ll}
\tilde x(\theta)=B \cos \theta,\\
\tilde y(\theta)=(B \sin \theta -A),\\
\end{array}
\right.
\end{equation}
so that the interface is the circle of center $(0,-A)$ and radius $B$.
Since the smoothed cusps are very narrow in practice ($\theta_0\ll 1$)
this circle arc is very flat; it tends to the straight horizontal
facet as $\theta_0\rightarrow0$.
The right end of this interface portion lies at
\begin{equation}
\left\{
\begin{array}{ll}
\tilde x_r=\tilde x(\pi/2-\theta_0)=B \sin \theta_0,\\
\tilde y_r=\tilde y(\pi/2-\theta_0)=B \cos \theta_0-A,\\
\end{array}
\right.
\end{equation}
where the local slope is
\begin{equation}
\tilde y_r'(\tilde x_r)=-\tan \theta_0 \label{Slope}.
\end{equation}
On the other hand, for $\theta \in [\pi/4,\pi/2-\theta_0]$
the crystal is bounded by the circle centered at point $P_c(\delta,\delta)$
(i.e. $x=y=\delta$) of rescaled radius unity, for which
\begin{equation}
\tilde R=(\tilde x_t-\tilde x_r)\sqrt 2=1-\sqrt 2 \sin \theta_0
\end{equation}
is a good first-order approximation when $\theta_0\ll 1$ 
(Fig.~\ref{fig:analeq}).

Note that, as $\theta_0\rightarrow 0$, one recovers the
equilibrium shape for the sharp cusp, with a horizontal facet
of length
\begin{equation}
\tilde \Lambda=2\tilde x_r\rightarrow 2\delta
\end{equation}
matching tangentially a circle of rescaled radius unity.

\begin{figure}
\centerline{
\psfig{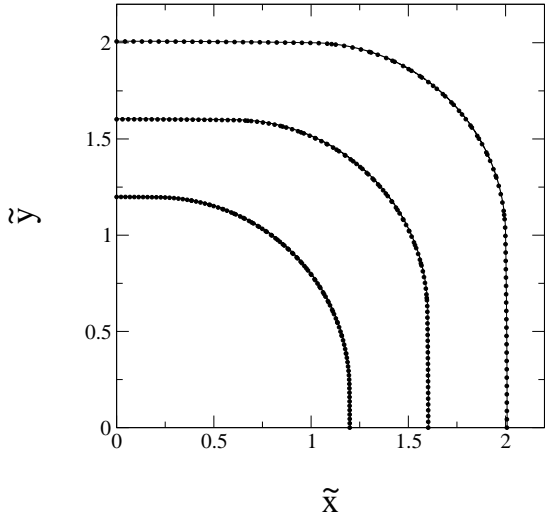}}
\smallskip
\caption{
Comparison of the analytical equilibrium shape (line)
with the phase-field solid-liquid interface (dots)
for $\delta=0.2,0.5,1.0$ (from left to right) and
$\theta_0=\frac{\pi}{200}$.
}
\label{fig:pfeq}
\end{figure}

As shown in Fig.~\ref{fig:pfeq}, the whole crystal shape is well reproduced
by the phase-field equilibrium code. A close examination of the numerical
data points reveals that the imposed anisotropy is underestimated
by about 0.2 percent in the numerics, independently of $\theta_0$.

\subsection{Numerical estimates for the facet ends}
\begin{figure}
\centerline{
\psfig{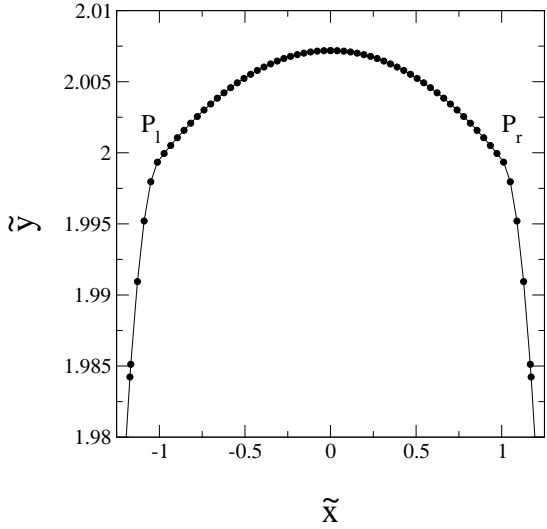}}
\smallskip
\caption{
Close-up of the phase-field equilibrium shape in the
region of the smooth facet ($\delta=1.0$, $\theta_0=\pi/200$).
Dots represent the interface points $P_i$.
}
\label{fig:faceteq}
\end{figure}

The knowledge of the analytical equilibrium shape
guides us to define a numerical procedure to extract estimates of
the facet length $\Lambda$ and corner
radius $R$ from discrete interfaces
obtained with the phase-field code.
On the square mesh, the interface consists in a list
of points $P_i(x_i, y_i)$. According to Eq.~(\ref{Slope}), the two 
facet ends are
located at the points $P_r$ and $P_l$ where the absolute value of the 
local slope
rapidly increases beyond $\tan \theta_0$
(Fig.~\ref{fig:faceteq}).
The method used to compute the two end points is thus to compare the 
derivative of
the equilibrium curve $\partial_x y$ with $\tan \theta_0$.
Using centered differences,
\begin{equation}
\partial_x y_i=\frac{y_{i+1}- y_{i-1}}{x_{i+1}-x_{i-1}}, \label{Centered}
\end {equation}
gives an error of about 5 percent on both measures.
On the other hand, using the one-sided approximations,
\begin{equation}
\partial_ x  y_i\vert^- =\frac{y_{i}-y_{i-1}}{x_{i}-x_{i-1}}, 
\label{Onesided_r}
\end {equation}
to compute $P_r$
and
\begin{equation}
\partial_x y_i\vert^+ =\frac{y_{i+1}-y_{i}}{x_{i+1}-x_{i}}, \label{Onesided_l}
\end {equation}
to compute $P_l$,
estimates of $\Lambda$ and $R$ are only 0.5 percent off the exact
values (see Table \ref{tab:1} ). This last estimation procedure is 
thus more precise and it is
readily extended to the case of growth shapes in what follows.

\subsection{Sharp cusp limit}
One may still wonder if the sharp cusp limit
($\theta_0\rightarrow 0$)
is reacheable within our numerical approach. To answer this question, it is
better to develop Eq.~(\ref{EqPF}) in terms of the second partial 
derivatives of $\psi$.
One gets then
\begin{eqnarray}
\tau(\theta)\partial_t\psi &=&
\big[\psi+\lambda \Delta(1-\psi^2)\big](1-\psi^2)\nonumber \\
&&+W_0^2\big[C_{xx}\partial_{xx}\psi
+C_{xy}\partial_{xy}\psi+C_{yy}\partial_{yy}\psi\big], \label{DevPF1}
\end{eqnarray}
with
\begin{equation}
C_{xx} = \left\{
\begin{array}{ll}
1+\delta(3\vert \cos \theta \vert - \vert \cos^3 \theta \vert+ \vert 
\sin^3 \theta \vert) +\delta^2,\\
A^2+B^2-AB \vert \cos \theta \vert (2+\sin^2 \theta),\\
B^2-AB \vert \sin^3 \theta \vert, \label{DevPF2}
\end{array}
\right.
\end{equation}

\begin{equation}
C_{xy} = \left\{
\begin{array}{ll}
2\delta(\sigma_s \cos^3 \theta+ \sigma_c \sin^3 \theta) + 2 \delta^2 
\sigma_s \sigma_c,\\
-2AB \sigma_c \sin^3 \theta,\\
-2AB \sigma_s \cos^3 \theta, \label{DevPF3}
\end{array}
\right.
\end{equation}
and
\begin{equation}
C_{yy} = \left\{
\begin{array}{ll}
1+\delta(3\vert \sin \theta \vert - \vert \sin^3 \theta \vert+ \vert 
\cos^3 \theta \vert) +\delta^2,\\
B^2-AB \vert \cos^3 \theta \vert,\\
A^2+B^2-AB \vert \sin \theta \vert (2+\cos^2 \theta), \label{DevPF4}
\end{array}
\right.
\end{equation}
for $\theta_0<\theta<{\pi\over 2}-\theta_0$, $\theta\le\theta_0$,
and $\theta\ge{\pi\over 2}-\theta_0$, respectively.
The symbols $\sigma_s$ and $\sigma_c$ represent the signs ($\pm 1$) of
$\sin \theta$ and $\cos \theta$.

The time step $\Delta t$ must be sufficiently small to ensure convergence
of the finite difference scheme, the most stringent constraint
on $\Delta t$ arising for $\theta=\theta_0$:
\begin{equation}
\Delta t \le  {\tau_0 \over W_0^2} {h^2 \over 2} \Big[{f_s^2(\theta_0)
\over C_{xx}(\theta_0)+C_{yy}(\theta_0)}\Big].
\end{equation}
As $\theta_0 \ll 1$, this approximates to
\begin{equation}
\Delta t \le {\tau_0 \over W_0^2} {h^2 \over 2} \Big({1+\delta \over 
\delta}\Big) \theta_0.
\end{equation}
It is thus always possible to reduce $\theta_0$ provided
that $\Delta t$ is decreased in proportion.
Although $\Delta t$ depends on the cusp amplitude $\delta$,
the time step can still be kept constant
while maintaining numerical stability by imposing
\begin{equation}
\theta_0=K{{\delta}\over{1+\delta}},
\end{equation}
with $K$ some constant. In the numerical simulations, we tipically 
take $K=\pi/100$,
so that $\Delta t/\tau_0 \le 0.0025\dots$. In practice, we rather use
the discretization scheme described in section III than the fully developped
version given in Eqs.~(\ref{DevPF1}-\ref{DevPF4}). The former proves 
to be more precise and
more stable than the latter, so that a larger time step, $\Delta 
t/\tau_0=0.008$,
can be used.

\section{Faceted needle growth}

\begin{figure}
\centerline{
\psfig{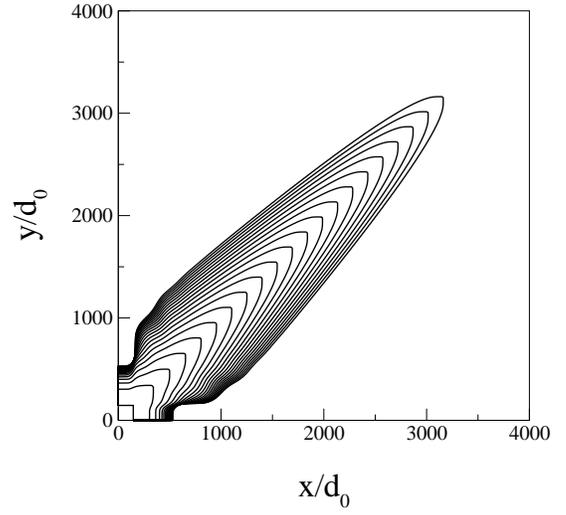}}
\smallskip
\caption{
Time evolution of a faceted dendrite
for undercooling $\Delta=0.55$, smoothing angle $\theta_0=\pi/200$,
and anisotropy $\delta=1.0$. The time interval between
two succesive curves is $100 \tau_0$.
}
\label{fig:pfevolutiona}
\end{figure}

We now turn to non-equilibrium growth shapes.
The time evolution of a needle dendrite is illustated in 
Fig.~\ref{fig:pfevolutiona}
for a cusp amplitude $\delta=1$,
diffusion coefficient $D\tau_0/W_0^2=4$ and undercooling $\Delta=0.55$.
The size of the simulation box is $600 W_0\times 600 W_0$.
For reasons of symmetry, it is sufficient to grow one half of the dendrite
(here $y\ge x$).
We checked that identical patterns are obtained when the
whole quadrant is used.
The initial conditions for the phase field are $\psi=1$ if
both $0\le x\le 20 W_0$ and $0\le y\le 20 W_0$
and $\psi=-1$ otherwise. This square shape is imposed
to the solid germ in order to get a
dendrite with a single tip along the $x=y$ direction. Conversely,
rounded initial shapes, like a circle, always result into
tip splitting along the growth direction, leading to more
complicated patterns which will be discussed later.
The temperature field $u$ is initially set to zero inside the solid germ
and to $-\Delta$ outside. Reflecting conditions
are imposed at all the domain boundaries

We first focus on the influence of the anisotropy
coefficient, $\delta$, on the dynamics of the
faceted needle dendrite. This coefficient is varied
from 0.25 to 1.60 in the simulations.
After a transient, the dendrite reaches a stationary state
which is independent of the size of the initial germ.
The tip then moves with a constant velocity $V$
and a rather large portion behind the tip adopts a stable shape.
A good test of the thin-interface limit
is to decrease independently the rescaled
interface thickness $W_0/d_0$ (or equivalently the diffusion coefficient
$D\tau_0/W_0^2$) and the grid spacing $h/W_0$ \cite{KarRap}.
Convergence is achieved when the scaled
tip velocity $\tilde V = V d_0/D$
is independent of both pameters.
Table \ref{tab:2} shows that
the results are reasonably well converged
for the values $D\tau_0/W_0^2=4$ and $h/W_0=0.4$
used in our simulations.

\subsection{Steady-state shape}
\begin{figure}
\centerline{
\psfig{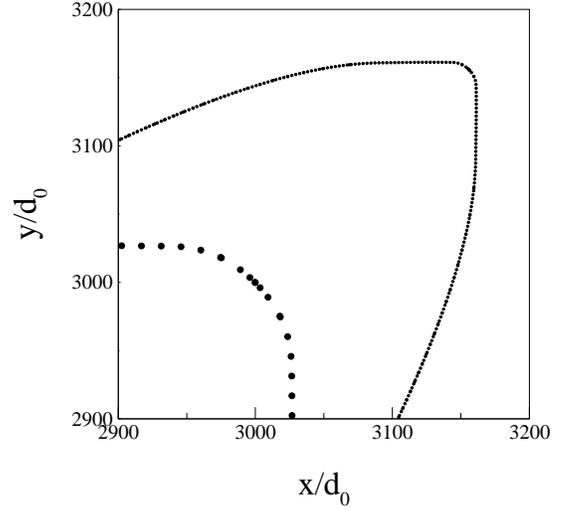}}
\smallskip
\caption{
Enlargment of the foremost dendrite tip
of Fig. \ref{fig:pfevolutiona}.
The tip is further enlarged
and shifted to the lower-left corner (larger dots).
}
\label{fig:pfevolutionb}
\end{figure}

\begin{figure}
\centerline{
\psfig{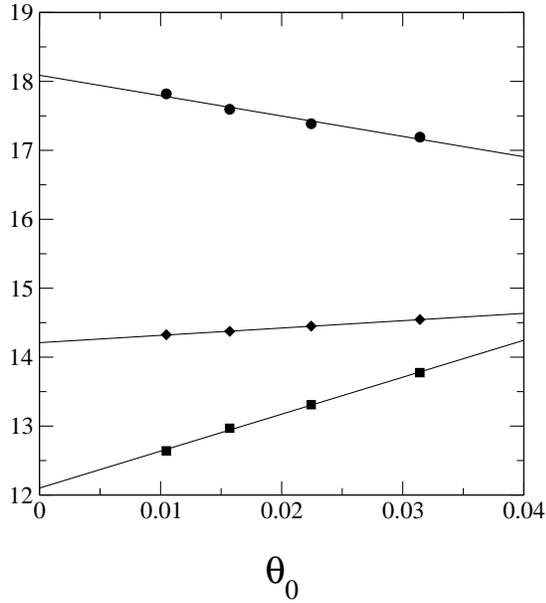}}
\smallskip
\caption{
Variations with the smoothing angle $\theta_0$ of
$R/d_0$ (circles), $0.25(\Lambda/d_0)$ (squares)
and  $50(V \tau_0/W_0)$ (diamonds) for anisotropy $\delta=1.0$.
The continuous lines are linear fits to the data points.
}
\label{fig:theta0}
\end{figure}

The steady-state interface shape can be divided into three
distinct parts: a nearly circular tip of radius $R$, a smooth
``facet'' of length $\Lambda$, and a trailing rough
tail (Fig.~\ref{fig:pfevolutionb}).
Estimates of $R=(x_t-x_r)\sqrt 2$ and $\Lambda=x_r-x_l$ are obtained
by using the numerical interpolation method described in section IV-B.
To smooth out temporal fluctuations, time
averages of both quantities are performed over the whole
stationary regime.
Obviously, the exact shape of the dendrite
tip should depend on the smoothing angle $\theta_0$.
When $\theta_0$ is sufficiently small this dependence is found to be
linear, which allows us to unambiguously
extrapolate $R(\theta_0)$ and $\Lambda(\theta_0)$
to $\theta_0=0$ (Fig.~\ref{fig:theta0}).
Up to an overall scale factor, the dendrite tip shape is very
similar to the equilibrium shape, i.e.
a quarter of a circle matching tangentially
to two side facets.

\begin{figure}
\centerline{
\psfig{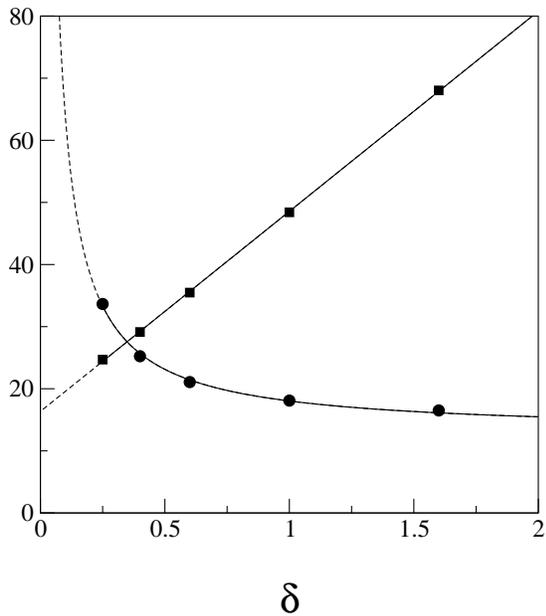}}
\smallskip
\caption{
Variations with the cusp amplitude $\delta$
of $R/d_0$ (circles) and $\Lambda/d_0$ (squares).
The continuous lines are least-squares
fits of the data points to the laws given in Eqs.~(\ref{Lambda}, \ref{Rtip}).
}
\label{fig:LambdaR}
\end{figure}

The variations of $\Lambda(0)$ and $R(0)$ with
cusp amplitude are dispayed in Fig.~\ref{fig:LambdaR}.
The results are well fitted by the two simple relations,
\begin{equation}
\Lambda=\Lambda_0\left[1+\frac{\delta}{\delta_\Lambda}\right],\label{Lambda}
\end{equation}
and
\begin{equation}
R=R_0\left[1+\frac{\delta_R}{\delta}\right].\label{Rtip}
\end{equation}
The asymptotic shape of the trailing rough part away from
the facet is well-fitted by a parabola that satisfies the
Ivantsov relation \cite{Iva47}
\begin{equation}
\Delta=\sqrt\pi \sqrt p \,e^p\, {\rm erfc} (\sqrt p) \label{Iva}
\end{equation}
with the P\'eclet number
\begin{equation}
p\equiv \frac{\rho V}{2D}.
\end{equation}
Note that the reason why the asymptotic tail is a
parabola whose tip radius is predicted exactly by the Ivantsov relation
is the same as for non-faceted growth \cite{Bre93,Kar:Houches}. Namely,
in a frame that is stationary with respect to the melt,
the trailing rough part grows as a planar interface whose position $x(t)$
perpendicular to the growth axis $\sim t^{1/2}$ and the
P\'eclet number $p\equiv x(t)(dx(t)/dt)/2D$ obeys the same
relation as Eq.~(\ref{Iva}), as shown independently
of Ivantsov by Zener \cite{Zen49}.

\begin{figure}
\centerline{
\psfig{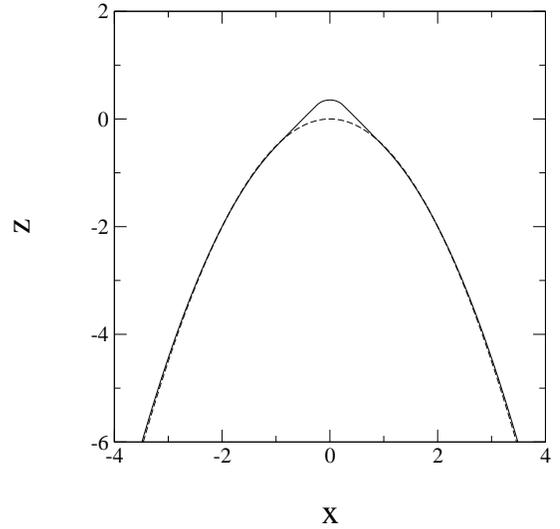}}
\smallskip
\caption{
Same needle dendrite as in Fig.~\ref{fig:pfevolutiona} after rotation,
translation, and normalization by $\rho=2Dp/V$ (solid line).
Also shown is the corresponding Ivantsov parabola $z=-x^2/2$ (dashed line).
Note that here, as in Fig.~\ref{fig:PFvsANneedle}, $z$ represents the
main axis of the dendrite, $x$ the axis perpendicular to $z$, and that
both $x$ and $z$ coordinates are normalized by $\rho$.
}
\label{fig:PfIvantsov}
\end{figure}

\begin{figure}
\centerline{
\psfig{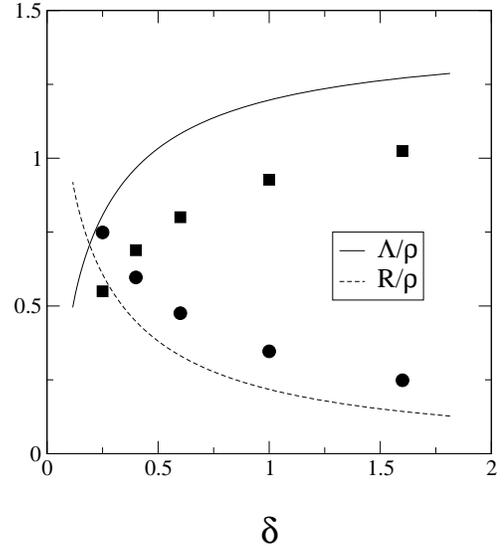}}
\smallskip
\caption{
Plots of $\Lambda/\rho$ and $R/\rho$ for $\Delta=0.55$,
obtained from phase-field simulations (squares and
circles, respectively) and predicted by the
sharp-interface theory of section VI (solid and dashed
lines, respectively).
}
\label{fig:ratios}
\end{figure}

We checked in a few simulations
that $\rho V$ is well predicted
by the Ivantsov relation, as illustrated in Fig.~\ref{fig:PfIvantsov}
where we compare the phase-field interface to the Ivantsov parabola
with the same tip velocity, i.e., the parabola with
$\rho=2Dp/V$, where $p$ is the Ivantsov P\'eclet
number and $V$ is the steady-state tip
velocity in the phase-field simulation. Given this agreement, we
use directly the formula $\rho=2Dp/V$ to calculate $\rho$ in all
results reported thereafter.
It should be emphasized the tip radius $\rho$ of this imaginary parabola
that matches exactly the asymptotic needle crystal shape and
the true tip radius $R$ are quite different as shown
in the plot of $R/\rho$ versus $\delta$ in Fig.~\ref{fig:ratios}.
In particular, $R/\rho$ is seen to decrease sharply with increasing $\delta$,
while $\Lambda/\rho$ approaches a constant. These plots reflect the fact that
side facets become elongated and extend closer to the tip that
becomes more pointed as $\delta$ increases.

\subsection{Steady-state operating state}
As in previous sharp-interface
calculations \cite{Vincent,Mokhtar},
we define the dimentionless tip selection parameter
\begin{equation}
C\equiv\frac{4\rho^2V}{D d_0},
\end{equation}
where $V$ is the tip velocity and $\rho$ is the
tip radius of the parabola that exactly matches the
asymptotic tail of the faceted needle crystal. Note that $C=8/\sigma$
where $\sigma=2Dd_0/\rho^2V$ is the selection parameter that is also
often used in dendrite growth studies.

\begin{figure}
\centerline{
\psfig{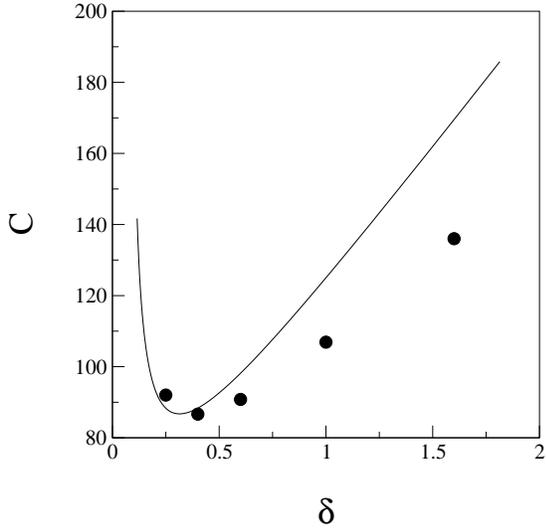}}
\smallskip
\caption{
Tip selection parameter $C=4\rho^2V/Dd_0$ versus
cusp amplitude $\delta$ for $\Delta=0.55$,
extracted from the phase-field simulations
(solid circles) and predicted by the approximate sharp-interface
analytical theory of section VI (solid-line).
}
\label{fig:Cvsdelta}
\end{figure}

Eliminating $\rho$ in favor of the P\'eclet number
using the relation $\rho=2Dp/V$, we can rewrite $C$ in
the form
\begin{equation}
C={16 D\over V d_0}p^2.
\end{equation}
A plot of $C$ versus $\delta$ with $V$ corresponding to the
steady-state tip velocity in the phase-field simulations
and $p$ computed from the Ivantsov relation for the value of
$\Delta$ corresponding to the simulations is shown in Fig.~\ref{fig:Cvsdelta}.
Note that $C(\delta)$ has a minimum that corresponds to
a maximum of tip velocity as a function of $\delta$.

\begin{figure}
\centerline{
\psfig{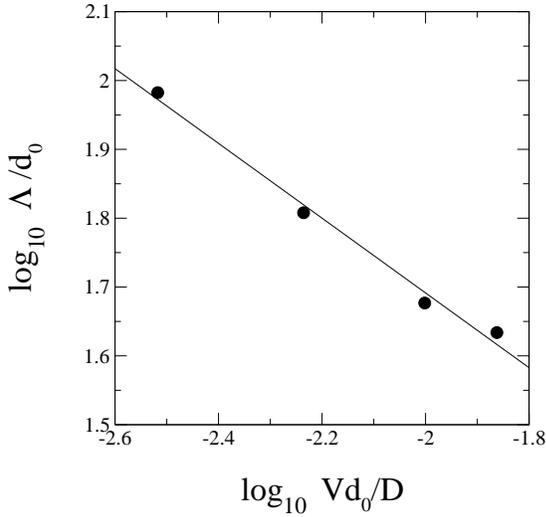}}
\smallskip
\caption{
Log-Log plot of the dimensionless facet length $\Lambda/d_0$ as a
function of the dimensionless tip velocity $Vd_0/D$ for a cusp amplitude $\delta=1.0$. The
straight line is a linear fit to the data points giving a slope $-0.54\pm0.04$.
}
\label{fig:logVlogLambda}
\end{figure}

In experiments, the growth rate of a single material
is usually studied as a function of undercooling
Additional simulations are thus performed
for other values of $\Delta$. All the other parameters are kept constant,
except for the total time which is increased to $7000 \tau_0$ for $\Delta=0.45$,
$3000 \tau_0$ for $\Delta=0.50$, and reduced to $1500 \tau_0$ for $\Delta=0.60$.
We do not extrapolate $V$ to $\theta_0\rightarrow 0$
here because the computing time becomes prohibitively
large for $\Delta<0.55$. However, from the previous results for
$\Delta=0.55$, we do not anticipate deviations in the tip velocity
by more than a few percent (see Fig.~\ref{fig:theta0}).
The plot of tip velocity versus facet length shown in 
Fig.~\ref{fig:logVlogLambda}
shows that our numerical results are consistent with the scaling law
$\Lambda\sim V^{-0.5}$ which was found experimentally for
NH$_4$Br needle crystals \cite{Maurer}.

\subsection{Transients}
\begin{figure}
\centerline{
\psfig{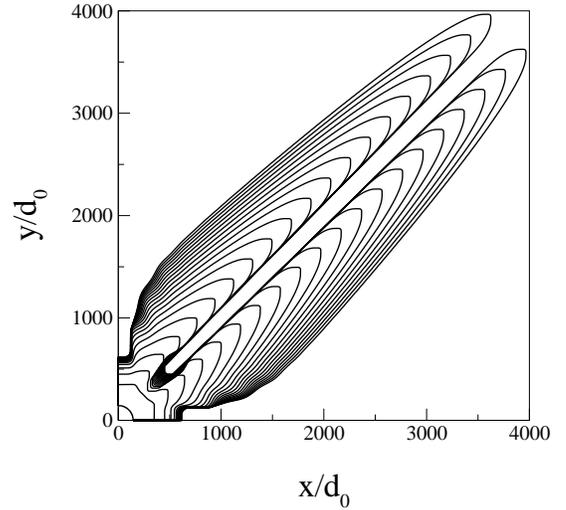}}
\smallskip
\caption{
Time evolution of a double needle dendrite
for undercooling $\Delta=0.60$ and anisotropy $\delta=1.0$
(time interval between two contours: 100$\tau_0$).
}
\label{fig:doubleneedle}
\end{figure}

When growth is started from a circular germ,
a double needle is systematically observed (Fig.~\ref{fig:doubleneedle}),
independently of the germ shape, for $\delta\le 0.20$.
Initially, the two tips move away from each other. Our simulations show that
these needles ultimately grow practically parallel to
each other at very long times, when they are separated
from each other by several diffusion lengths
(Fig.~\ref{fig:doubleneedle}).

\begin{figure}
\centerline{
\psfig{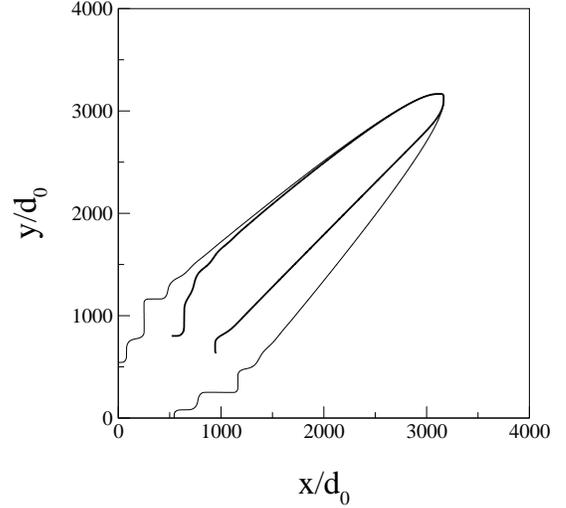}}
\smallskip
\caption{
Comparison between one branch of a double needle
dendrite (thick line) and  a single
needle dendrite (thin line). The first
curve is shifted to superimpose the two tips.
Time $t=1500\tau_0$, undercooling $\Delta=0.60$
and anisotropy $\delta=1.0$.
}
\label{fig:DoubleSingle}
\end{figure}

\begin{figure}
\centerline{
\psfig{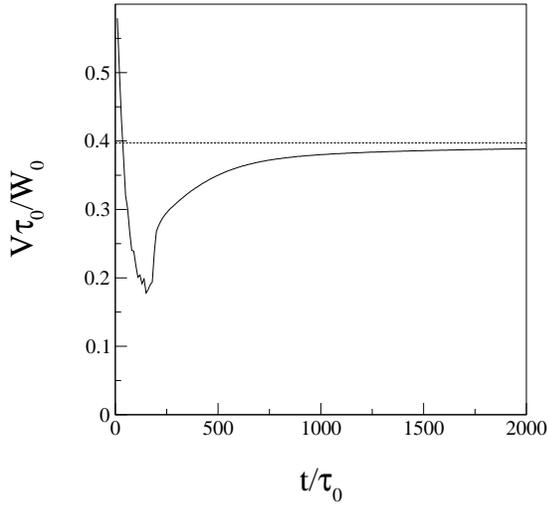}}
\smallskip
\caption{
Tip velocity as a function of time for the double needle dendrite
shown in Fig.~\ref{fig:doubleneedle}. The dashed horizontal line 
gives the stationary velocity
obtained for the single needle dendrite with the same parameters.
}
\label{fig:Vdouble}
\end{figure}

At long times, the tip of the double needle becomes comparable
in shape to that of the single needle dendrite (Fig.~\ref{fig:DoubleSingle}).
Of course, the former is thinner on one side because
the latent heat accumulates in the channel between the two tips.
This similarity in shape suggests that the double needle pattern is governed by
the same operating state as the needle dendrite. Verifying
this point with a sufficient accuracy is not easy because of a still
longer transient for the double needle dendrite. For this reason, we
compare the patterns obtained for a larger undercooling, $\Delta=0.60$,
for which the simulations are still quantitative
but the relaxation time much shorter than with $\Delta=0.55$. After
a time $t=2000\tau_0$, the tip velocity of the double needle pattern has
not yet fully converged but it tends slowly to a limit
near $0.397W_0/\tau_0$ (Fig.~\ref{fig:Vdouble}).
This is precisely the value found for the single tip dendrite, which confirms
that the two asymptotic states are identical. Thus, the double needle dendrite
is merely just a tip-splitting evolution of the single tip dendrite.
This splitting is often preferred by the system because it
allows for a more efficient occupation of the available room.

Let us finally remark that the double needle found here is not a
faceted version of the generic non-faceted growth
structure that has been termed doublon \cite{Ihle&Muller}.
Doublons also are symmetry-broken growth shapes with a double tip. However,
the two tips remain much closer than one diffusion length
and are thus highly inter-dependent, which is not true for
the double faceted needle dendrite observed in the present simulations.

\section{Comparison with sharp-interface analytical theory}

A precise quantitative comparison of the phase-field
results of the last section with previous sharp-interface
calculations \cite{Vincent,Mokhtar} cannot be carried out
because these calculations considered the one-sided model with zero
diffusivity in the solid-phase while we simulated here a symmetric
model with equal diffusivities in solid and liquid. In addition,
our simulations are for finite P\'eclet number while the sharp-interface
calculations of Refs. \cite{Vincent,Mokhtar} are restricted
to the small P\'eclet number limit ($p\rightarrow 0$).
Nonetheless, the phase-field results
reproduce several key features of faceted needle
growth that were predicted in the sharp-interface studies:
\begin{enumerate}
\item
The ratio $\Lambda/\rho$ of the facet length to the tip
radius of the parabola that matches the asymptotic
rough tail increases with the cusp amplitude $\delta$ and
saturates to a value of order unity
for large $\delta$.
\item
The tip selection constant $C$
has a minimum for a value of $\delta$ of order unity.
The value of $C$ that corresponds to this minimum
is large ($\sim 100$).
\item
For large $\delta$, the needle shape consists of two long facets joined
by a nearly circular tip that match on to trailing rough parts
that become parabolic far from the tip. This is precisely
the shape proposed by Adda Bedia and Hakim \cite{Vincent}.
\end{enumerate}

In the remainder of this section, we develop a simple analytical
theory of faceted needle growth that is based on the approximate
shape proposed by Adda Bedia and Hakim \cite{Vincent} for large
$\delta$, where facets are joined by a small circular tip of radius $R$.
We take this approximation further by assuming that (i) it remains
valid for $\delta$ of order unity, and (ii) the trailing rough parts remain
parabolic all the way to the points at which they match tangentially
the facets, as opposed to being parabolic
only asymptotically far from the tip \cite{Vincent}.
These simplifications allow us to obtain a simple physical picture
of faceted needle growth as well as explicit analytical predictions 
of the shape
parameters $R$, $\Lambda$ and $\rho$ and the velocity
$V$ for arbitrary P\'eclet number, without necessitating the
numerical solution of an integral equation as in Refs.
\cite{Vincent,Mokhtar}. As we shall see below, these predictions
agree reasonably well with our phase-field results despite our
simple parametrization of the steady-state shape.
 
\begin{figure}
\centerline{
\psfig{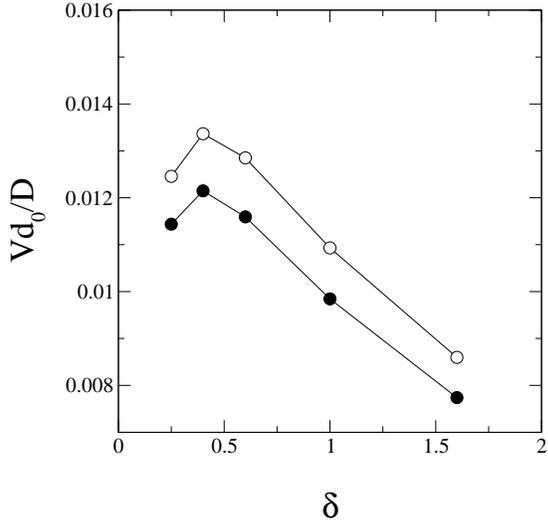}}
\smallskip
\caption{
Variation of dimensionless tip velocity $Vd_0/D$ with
cusp amplitude $\delta$ obtained from the phase-field
simulations for $\Delta=0.55$ (solid circles)
and using the equation $V=2Dp/\rho$, where $p(\Delta)$ is
the P\'eclet number predicted by the Ivantsov relation, and
$\rho=(\Lambda+R)/ \sqrt 2$, where $\Lambda$ and $R$ are the facet
length and tip radius obtained from the
phase-field shapes (empty circles).
}
\label{fig:vtilde}
\end{figure}

The complete determination
of the needle shape ($R$, $\Lambda$ and $\rho$)
and the growth velocity $V$ for fixed $\Delta$ and
$\delta$ requires four independent
relations. The first is the Ivantsov relation (\ref{Iva})
that fixes the product $\rho V$. A second relation between
$R$, $\Lambda$ and $\rho$ is simply obtained by imposing that
the circular tip and parabolic tails match tangentially
the front and trailing ends of the facets, respectively, which yields
at the relation
\begin{equation}
\rho=(\Lambda+R)/\sqrt 2.  \label{shape} 
\end{equation}
We can check this relation by comparing
the tip velocity in the phase-field simulations
with the velocity
$V=2Dp/\rho$, where $p$ is the P\'eclet number
predicted by the Ivantsov relation  and $\rho$ is computed using
Eq.~(\ref{shape}) together with values of $\Lambda$ and $R$
in the simulations.
Fig.~\ref{fig:vtilde} shows that these two velocities
are in reasonably good quantitative agreement.
The systematic deviation is most likely due
to the fact that the crystal tail is
not exactly parabolic very close to the facet.

\begin{figure}
\centerline{
\psfig{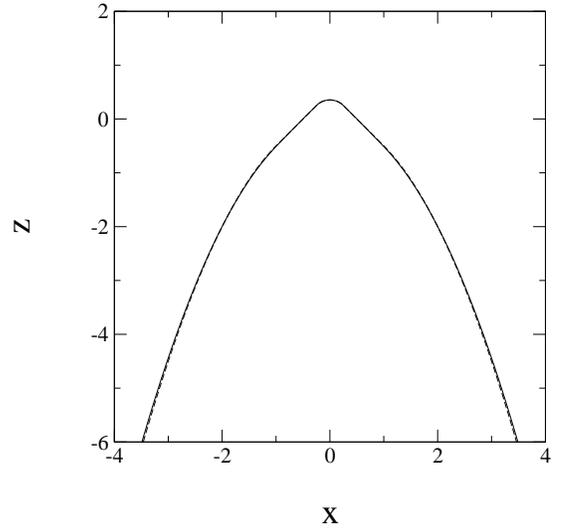}}
\smallskip
\caption{
Comparison of the phase-field needle dendrite of Fig.~\ref{fig:PfIvantsov}
(solid line) with the faceted needle crystal shape
assumed in the analytical theory with a circular tip
and a parabolic tail (dashed line). Both curves are for
the same parameter value $R/(\sqrt 2 \rho)=0.25$.
}
\label{fig:PFvsANneedle}
\end{figure}

The two other relations needed to complete our
theory are obtained by imposing the
Gibbs-Thomson condition (\ref{gt}) at the tip of the needle
crystal and by using the integral form of this condition
on the facet (Eq.~\ref{int}).
For this, we need an expression for the
undercooling along the interface that is obtained
straightforwardly using the known
boundary integral relation \cite{NashGlick74,BarLan89}
\begin{eqnarray}
& &u(x)=-\Delta+p\int_{-\infty}^{+\infty}  \frac{dx'}{\pi}
\exp\left(-p\,[z(x)-z(x')] \right)\nonumber \\
& & \times K_0\left(p\sqrt{(x-x')^2+(z(x)-z(x'))^2}\right),\label{green}
\end{eqnarray}
where $K_0$ is the zeroth order modified Bessel function, $p$ is
given by the Ivantsov relation (\ref{Iva}),
and $z(x)$ is the interface shape with length in unit of $\rho$.
For a circular tip and parabolic tails matching tangentially
side facets that make
a $45^o$ angle with respect to
the growth axis, the $x$ coordinates of the front and trailing ends 
of the facets are
$x=\pm R/\sqrt{2}\rho$ and $x=\pm 1$, and
\begin{eqnarray}
& & z=
\left(\frac{R^2}{\rho^2}- x^2\right)^{1/2}- \frac{\sqrt{2}R}{\rho}+
\frac{1}{2},~~~~0\le| x|\le \frac{R}{\sqrt{2}\rho},\label{ztip}\\
& &  z=-| x|+\frac{1}{2},~~~~
\frac{R}{\sqrt{2}\rho }\le| x|\le 1, \label{zfac}\\
& &  z=-\frac{ x^2}{2},
~~~~ | x|\ge 1. \label{ztail}
\end{eqnarray}
These equations describe rather accurately
the phase-field needle dendrites (Fig.~\ref{fig:PFvsANneedle}).
Since the isothermal
Ivantsov parabola is an exact solution
of the steady-state growth problem without capillarity,
the integral relation (\ref{green}) is exactly
satisfied for $z=-x^2/2$ and $u=0$. In contrast,
as illustrated in Fig.~\ref{fig:PfIvantsov},
the tip of the faceted needle crystal protrudes ahead of the tip of this
parabola and hence is undercooled by a finite amount $u(0)$.
The Gibbs-Thomson condition (\ref{gt}) imposes a relation between
this tip undercooling and the tip radius that is simply
\begin{equation}
u(0)=-d_0/R. \label{gttip}
\end{equation}
In turn, Eq.~(\ref{int}) provides us with a
relation between the average undercooling on the facet
and the cusp amplitude $\delta$ that takes the form here
\begin{equation}
\sqrt{2}\int_{\frac{R}{\sqrt{2}\rho}}^{1}\,dx\,u(x)=-2\frac{d_0}{\rho}\,\delta,
\label{avgu}
\end{equation}
where $x$ is in units of $\rho$ as above. The four relations
defined by Eqs.~(\ref{Iva}, \ref{shape}, \ref{gttip}) and (\ref{avgu}),
together with the expression for $u$ on the interface defined by
Eqs.~(\ref{green}-\ref{ztail}), completely determine
$R$, $\Lambda$, $\rho$ and $V$. Since the product $\rho V$ is
exactly predicted by the Ivantsov relation, we only need to
compute three independent dimensionless combinations
of the four above quantities to compare the
predictions of the above theory with the phase-field
results. The most meaningful dimensionless combinations
are the selection parameter $C$ and the two ratios $\Lambda/\rho$ and
$R/\rho$. To compute those as a function of $\delta$, it is
convenient to vary $R/\rho$ and compute $C$, $\delta$, and
$\Lambda/\rho$ using the following relations
that are simple to deduce:
\begin{eqnarray}
& &\frac{1}{C}=-\frac{\sqrt{2}}{16 p \,
\delta}\int_{\frac{R}{\sqrt{2}\rho}}^{1}\,dx\,u(x),\\
& &\delta= \frac{\rho}{\sqrt{2}u(0)R}\int_{\frac{R}{\sqrt{2}\rho}}^{1}\,dx\,u(x), \\
& &\frac{\Lambda}{\rho 
}=\sqrt{2}\left(1-\frac{R}{\sqrt{2}\rho}\right).
\end{eqnarray}
The above predictions are compared
with the phase-field results in
Figs.~\ref{fig:ratios} and \ref{fig:Cvsdelta}. The
quantitative agreement between the two is reasonably
good but not exact because the steady-state shape
that we have assumed here differs slightly from the true shape
(Fig.~\ref{fig:PFvsANneedle}) that is only exactly parabolic 
asymptotically far from the facet.

To conclude, we note that the expression for $u(x)$ can be simplified
in the small P\'eclet number limit where Eq.~(\ref{green})
reduces to \cite{PelPom86}
\begin{equation}
u(x)=-p\int_{-\infty}^{+\infty}  \frac{dx'}{2\pi}
\ln \left[\frac{(x-x')^2+(z(x)-z(x'))^2}{(x-x')^2+(x'^2/2-x^2/2)^2}
\right].
\end{equation}
Therefore, both $u(0)$ and the integral of $u(x)$ along the facet are simply
proportional to the P\'eclet number in this limit and hence
$C$, $R/\rho$, and $\Lambda/\rho$ are only functions of $\delta$.
Furthermore, it is straightforward to deduce that $\Lambda/\rho\rightarrow
\sqrt{2}$, $R/\rho\rightarrow 0$, and $C\sim \delta$ in the large
$\delta$ limit. Hence this theory predicts
the scaling law $\Lambda\sim V^{-1/2}$ that is a simple
consequence of the fact that $\Lambda/\rho$ and $C\sim \rho^2V$ are constants
in the small P\'eclet number limit.

\section{Concluding Remarks}

In conclusion, we have shown that the phase-field approach can
be successfully extended to model the solidification of faceted
materials. Our approach, which
consists of rounding the cusps in the $\gamma$-plot,
converges well in the limit of sharp cusps for both 
equilibrium and non-equilibrium growth shapes with facets. 
Even though we have
considered a simple form of the $\gamma$-plot, 
the method should be applicable to more complex $\gamma$-plots where
the interface stiffness varies on the rough parts.

In addition, we have developed an approximate analytical
theory of faceted needle growth that includes capillarity and
assumes circular and parabolic forms for the front and
trailing rough parts of steady-state needle crystals, respectively.
This theory yields explicit predictions of tip velocity
and facet length that are in good overall quantitative agreement
with the phase-field results; this agreement is largely due to the 
fact that the needle shape assumed in the theory 
is a very good approximation to the actual growth 
shape observed in the phase-field simulations. Furthermore, this theory
predicts that the scaling law $\Lambda\sim V^{-1/2}$ observed both
experimentally \cite{Maurer} and in the present phase-field simulations
should hold in the small P\'eclet number limit,
consistent with the results of previous 
theoretical studies \cite{BenPom88,Vincent,Mokhtar}.

In the present study, we have established a quantitative
comparison between phase-field and sharp-interface results
under the assumption that the variation of interface undercooling is
dominated by capillarity. 
Using scaling arguments,
Ben Amar and Pomeau \cite{BenPom88} have concluded that, under the assumption
that the growth of facets is dominated by a Franck-Read
screw discolation mechanism, the kinetic
undercooling of the interface should be negligible
over an intermediate range of velocity where the $\Lambda\sim V^{-1/2}$ 
scaling law has been observed \cite{Maurer}. While growth conditions where kinetic
effects are negligible may exist, we suspect that such effects
will generally be important in the presence of facets, as widely
believed. Therefore, a quantitative incorporation of facet kinetics in a 
phase-field model remains an important task for the
future. The phase-field model as formulated thus far
reproduces a linear relationship between the planar interface velocity and
interface undercooling appropriate for a rough interface. On facets, however, 
standard growth mechanisms such as screw dislocations or ledge nucleation
lead to a non-linear relationship between velocity and undercooling. These
relationships could potentially be incorporated in the phase-field model by letting
the kinetic relaxation time $\tau$ depend on temperature or supersaturation,
in addition to orientation. 

Another interesting future prospect is 
to model the directional solidification of alloys with faceted
interfaces \cite{Trivedi,Salol,Henry}
by combining the present methodology to handle cusped $\gamma$-plots 
with a recent thin-interface phase-field formulation
of alloy solidification  \cite{Kar01}.

\acknowledgments {
A.K. wishes to thank the hospitality of the
Laboratoire Mat\'eriaux et Micro\'electronique de Provence,
Universit\'e d'Aix-marseille III where this work was initiated
as well as the support of U.S. DOE Grant
No. DE-FG02-92ER45471. We also
thank Vincent Hakim and Klaus Kassner for valuable discussions.
This research benefited from computer time
allocation at the Advanced Scientific Computation Center
at Northeastern University.
}

\vskip50pt

\noindent{\bf Tables:}

\begin{table}[h]
\caption{Comparison of the facet length (first column)
and corner radius (second column)
of the equilibrium shape for $\delta=1.0$ and $\theta_0=\pi/200$.
Analytical values are compared with estimates from phase-field
data analyzed with two different interpolation methods.}
\begin{tabular}{ccccc}
$\tilde x_r-\tilde x_l$&$(\tilde x_t-\tilde x_r)\sqrt 2$&\\
\tableline
1.950&1.034&centered diff. approx. [Eq.~(\ref{Centered})]\\
2.023&0.983&one-sided approx. [Eqs.~(\ref{Onesided_r},\ref{Onesided_l})]\\
2.031&0.978&analytical\\ 
\end{tabular}
\label{tab:1}
\end{table}

\begin{table}[h]
\caption{Convergence of the steady-state tip velocity with decreasing
ratio $W_0/d_0=(D\tau_0)/(a_1a_2W_0^2)$
and lattice parameter $h/W_0$. The other parameters
are kept constant, $\Delta=0.55$,
$\delta=1.0$, $\theta_0=\pi/200$, and the kinetic coefficient is set to zero.}
\begin{tabular}{cccccc}
$D\tau_0/W_0^2$&$W_0/d_0$&$h/W_0$&$\Delta t/\tau_0$&$V\tau_0/W_0$&$Vd_0/D$\\
\tableline

3&5.42&0.4&0.008&0.158&0.0097\\ 
4&7.22&$-$&$-$&0.285&0.0098\\
5&9.03&$-$&$-$&0.402&0.0089\\
4&7.22&0.2&0.002&0.285&0.0098\\
$-$&-&0.4&0.008&0.285&0.0098\\
$-$&-&0.6&0.018&0.284&0.0098\\
$-$&-&0.8&0.032&0.280&0.0096\\
$-$&-&1.0&0.050&0.275&0.0094\\
\end{tabular}
\label{tab:2}
\end{table}

\end{document}